# Geolocation of an aircraft using image registration coupling modes for autonomous navigation

Nima ZIAEI

*Abstract*—This paper proposes to study an alternative technology to the GPS system on fixed-wing aircraft using the aerial shots of landscapes from a ventral monocular camera integrated into the aircraft and based on the technology of image registration for aircraft geolocation purpose. Different types of use of the image registration technology exist: the relative registration and the absolute registration. The relative one is able to readjust position of the aircraft from two successive aerial shots by knowing the aircraft's position of image 1 and the overlap between the two images. The absolute registration compare a real time aerial shot with pre-referenced images stored in a database and permit the geolocation of the aircraft in comparing aerial shot with images of the database. Each kind of image registration technology has its own flaw preventing it to be used alone for aircraft geolocation. This study proposes to evaluate, according to different physical parameters (aircraft 'speed, flight altitude, density of image points of interest ...), the coupling of these different types of image registration. Finally, this study also aims to quantify some image registration performances, particularly its execution time or its drift.

## I. Introduction

Many companies develop aircraft aimed to intervene on a broad spectrum of missions such as the surveillance of territory, the non-exposure of the pilots in hostile environment, the remote assistance to the persons in danger. One of this kind of aircraft is the Argodrone developed by Tyrix Aerospace Company.

To be certain of achieving these objectives, this 5m-wide thermal drone is a concentrate of technical prowess: its patented system of curly wings to lower its stall speed is a good illustration.

This aircraft by its characteristics belongs to the category of F civil aircraft, which circulate without any person on board. It can currently fly over French territory only subject to the establishment of an airworthiness document in accordance with the decree of 11 April 2012 on the free movement of civil drones [1], which classifies the aircraft and the part of the factor human being in the piloting of an **F** class civil drone. The best way to guarantee an airworthiness document is to ensure an optimal level of security, which is characterized by redundant systems. The Argodrone navigates by GPS. GPS is the most widespread system in the world for geolocation because it is relatively accurate in its measurement and has good coverage in terms of signals. We want to study an alternative technology to this geolocation system, to ensure the redundancy of the navigation system. The other existing geolocation systems mainly use an inertial measurement unit that has a significant drift. It is mainly due to measurement inaccuracies of the instrumentations but also to physical effects not taken into account such as the drift of the Earth's gravity as a function of the altitude and position of the system itself, or the phenomena of friction of the inertial instruments in the system. Consequently, the navigation by inertial unit requires being resetting by the conventional GPS system with a frequency whose value depend on the price of the technology (several hundred thousand US dollars [2]).

Naturally, our choice turned to a method of geolocation by image registration which offers an interesting alternative in terms of embedded solution for a given aircraft and at low cost because the technology is only using a monocular camera to make the shooting. The principle of geolocation by images registration lies in the shooting at regular intervals of landscapes overflown during the flight of the aircraft. The result of this shot is a picture called the current image. Prior to the flight, a DataBase (DB) is created with all the images of the landscapes that the aircraft will fly over, each image being geo located. The goal is to find the current image in the DB, using only image processing techniques. To find the current image among the images of the DB is to make an absolute space registration, that is to say to use a pre-geo image database located in order to calibrate the geographical position of the center of our current image. The absolute space registration technique makes it possible to know the position that the aircraft had at the time of the shooting of the current image. Thus, the absolute registration technique remains reliable in terms of accuracy by its nature (a calibration is always an exact measurement). However, the operational context in which we have to use the absolute space registration, would not allow us to use it only as the only technique possible in view of its potentially long calculation time and the constraints of the embedded. Thus, several studies [5], [6], [7], [12], [13] have

This work was entirely supported by the company ALTEN whose head office is located at 40 Avenue André Morizet, 92100 Boulogne-Billancourt and its stimulating innovation center located in the city of Chaville, FRANCE.
N. ZIAEI; is with the stimulating innovation center of ALTEN Company.

N. ZIAEI e-mail contact : nima.ziaei@alten.com

demonstrated that for an image registration system to be reliable, it is also necessary to introduce the technique of relative registration of images. This technique uses the regular shooting of consecutive current images having a common part, to allow the system to avoid systematic use of DB and to consume a large calculation time. Moreover, these same studies have concluded that it is impossible to use only the relative registration technique to perform geolocation, given its cumulative drift, which can become very important in the field time, same as for the case of an inertial measurement unit. It is therefore important to study a hybrid solution of absolute and relative registration image techniques to allow the aircraft to geo locate autonomously.

The registration itself, whether absolute or relative, uses image descriptor algorithms that represent the physical characterization of points of interest of an image, i.e., an imprint of the topographic details of an image. Describing an image via image descriptors makes it possible to lighten the image storage DB of the aircraft but also, to accelerate the identification time of a topographic scene by avoiding working pixel by pixel an HD image, but only through its descriptors of points of interest. If it is necessary to use the coupling between the two techniques of image registration, their coupling mode remains to be determined. Indeed, if the absolute registration serves to recalibrate the cumulative unit drift system of the relative registration, there is no point in being used systematically because the aircraft must know its position at a reasonable frequency, something that is not guaranteed with the only absolute adjustment. On the other hand, if absolute readjustment is used infrequently, the significant cumulative drift could make geolocation of the aircraft compromising and would require more DB images to be exploited when the system performs the absolute registration, which would make Absolute registration long to execute and of itself, would create a more important drift. The article proposes to study this compromise through a mathematical modeling approach whose some parameters have to be evaluated experimentally.

### A. *Description of our objectives*

As will be described in the state of the art of this article, many techniques of absolute and / or relative registration image in various fields (navigation, cartography, augmented reality) exist but no study makes it possible to study the precise performance of how these two techniques work once coupled. As mentioned in the introduction to this article, the purpose of this article is precisely to study the entanglement of the two relative and absolute image registration. The type of coupling of the two techniques will be called the image registration operating modes. The main criterion of image registration's operating modes 'success is the capacity to be able to do the most accurate geolocation of an aircraft, independently of a GPS system; the physical parameters (aircraft's speed , altitude of flight, optics field of view, ...) are potentially playing on the performance of each mode. This article proposes a modeling of different operating image registration modes and a study on their limits of exploitation. In practical terms, the article also proposes the mathematical tools exploiting the image descriptors to evaluate the movement of the aircraft by relative image registration. For this purposes, we assume that the aircraft remains approximately at the same altitude or is equipped with a system able to make image rectification during the entire maneuver. This assumption is reasonable given the performance of a fixed-wing aircraft to maintain any flight plan.

### B. *Design approach*

Given the objectives of the study, the first part of this article requires introducing one by one the possible operating modes in way of coupling the two image registration techniques. The second part will quantify some important physical features occurring in the performances of relative and absolute registration image's techniques.

This article is organized as follows: Section II examines the state of the art, Section III presents the main physical parameters at the origin of the degradation of the registration performance, and Section IV models the different scenarios of the coupling modes between the two types of image registration. The section V gives some theoretical answers to the practical use of the image registration in providing some results in the performances of the relative's one in the context of taking aerial images of urban topography. Finally, the section VI will conclude the study.

## II. KNOWLEDGE REPORT ON THIS FIELD

As in any innovative technology, the military domain predominates over advances in geolocation techniques and more precisely by using image registration for geolocation. To navigate, military technologies have a real need for redundant navigation systems that can bypass GPS technology. This is where the best performance is achieved and where the most interesting studies are done.

### A. *Images registration application to geolocation*

The article [3] describes the guidance of a submarine by resetting the reconstructed images of the seabed by merging data from its sonar system. The conditions of use are too far from those of an aircraft, making these results unworkable. The guidance of a satellite when it is put into orbit also uses image registration technology (article [4]). However, the conditions are also different in terms of speed and altitude. In addition, the technology used is much more efficient than the one we plan to use. Finally, the TERCOM system allows the guidance of a missile (article [5]), in various conditions, some similar to ours. We also have a substantial difference between the material means envisaged on the aircraft and those used by the army on a missile. The existence of this technology, however, demonstrates that the registration between a shooting and a pre-geo-localized image in a database is realistic and this technique is studied in our study under the name of absolute image registration. It offers an accuracy ranging from 30m to 90m. The implemented method processes the data of the database but also the data of the shooting image and finally carries out the image registration by correlation of these two types of information. We thus note that the pretreatment aspect is essential to realize the absolute registration of the images. This publication also mentions the DSMAC system, which makes it possible to identify the target and to realize its approach more precisely by relative image registration. The two methods of image registration (absolute and relative) are thus used to guide the missile: TERCOM to achieve the approach of the target, and DSMAC to reach it. Tomahawk missiles are already using this technology.

*B. Image registration application to cartography*

Image registration is used to create 3D maps, as explained by Royer [6] who designed an embedded system whose primary purpose is to map an urban area in 3D. This system does not meet our need for the operational context, but gives an example of building a database. However, we cannot restrict our aircraft to having to "explore" and build its own database as it goes. The move cannot be groped. The map registration on satellite using shot images [7] also makes use of image registration, exploiting only the high-level primitives (urban area and intersection). This article does not answer our problem however; it shows that some image details (those taken at high altitudes) are more reliable than others (at low altitudes) are.

*C. Images registration application to a location of an object in a landscape*

The article [8] presents a drone using a database on the ground, whose goal is to follow a target on the ground. Although the approach is detailed, we do not have enough precision on the algorithms used to determine if the study is an interesting solution for us. Article [9] gives an example of geolocation of a land vehicle from an air vehicle. Our technology is to geo locate the air vehicle and not necessarily a particular point on the ground. This technique is therefore not applicable for our study. The study of article [10] has developed a solution for tracking vehicles on a battlefield from a flying device. They exploit SAR (Synthetic Aperture Radar) images and a database of geo-referenced images. A base on the ground performs the treatments, in order to have sufficient computing power (parallel processing ...). This study is confidential, so we cannot access the technical details it contains. Han et al. [11] have developed a solution to perform target tracking in known ground, however the accuracy developed is several tens of meters away, so too far from our goal of providing redundancy with the GPS system, and more technical details of the project are not available.

*D. Aerial image registration*

Paper [11] proposes a rather different use of solutions presented in this study. Their solution is based solely on an estimate of the position from aerial images. The main goal is to develop a real-time implementation based on computer vision. First, the two types of geolocation by image registration are described. It is shown that these two techniques are complementary. Relative image registration provides fast information to calculate and does not depend on a database. However, the drift is cumulative. It means that the error in geolocation increases with time. It is therefore not able to ensure reliable relocation over time. However, by coupling this solution to the absolute registration, the error is regularly corrected. Thus, the information provided remains relatively reliable. The document highlights the following elements:
- The movements of the aircraft around its roll axis are the ones that induce the most error (distortion of the image), even for a small variation.
- The previous point has implied the following hypothesis: the axis of the camera must remain perpendicular to the tangent to the ground.
- The camera that captures images in the overhead area captures images with a frequency of 1 Hz
- Absolute resetting is much more efficient over areas of high contrast density (road networks, city) than in poor contrast ones (forests, mountains).

This interesting solution offers an option that has not been considered so far, namely the relative registration. It is a less reliable solution than absolute registration but offers many advantages, including independence from a database.

Paper [13] also uses this solution. Namely, a relative image registration carried out by means of an odometer. The characteristics are calculated on each image $n$ and match with those of the image $n + 1$. The association is established by minimizing the sum of the absolute differences squared. The smaller the difference in motion between two images, the more efficient is the algorithm. It therefore requires a high refresh rate (20-30Hz)

Paper [14] offers a solution to geo locate moving targets from a device in flight. The fact that the geolocation of the device is known involves a major difference with our subject but the article still gives us some interesting information. The order of magnitude of accuracy on this type of system is of the order of ten meters for an altitude ranging from 100 to 200m.

*E. This paper's innovation*

Images registration is currently used in many applications: **navigation, mapping or target tracking.** Several other fields use this technique: augmented reality, medical imaging (rigid and multimodal). Systems evolve either in a different operational context or require too much computing capacity. Many technical solutions are not sufficiently documented, or are confidential. The article [5] shows us that the resetting works at a speed of the order of 800km/h, is based on the absolute resetting for the trip over a long distance, and passes to the relative registration to have the precision required to reach his target. Even if we do not have the technical details, especially on the algorithms that allow to hybridize the two modes of registration, in a predictable way, this gives us an example of a functioning and existing double registration system. Our system will have to be both a navigation system that can be exploited at high speed (over 100km/h) and have an acceptable accuracy for localization. The realization of this compromise will constitute the benefit of our production compared to pre-existing applications. In addition, several applications use a database, or a remote calculation system. We exclude this solution because it does not allow ensuring the redundancy of GPS and autonomy of the aircraft: having to maintain a connection between the aircarft and a base is a risk factor that we want to be freed. The constraint in terms of weight of our aircraft is less critical (device of 150Kg) which makes it possible to board the material necessary for a database and an on-board computer. These different articles allow us to approximate the use made of geolocation by image registration at present as well as to know what are and where are the existing techniques.

However, we can also conclude that there is currently no prototype capable of achieving the appropriate type of coupling between absolute and relative image registration solely by knowing the topography overflown and knowing the flight parameters of the aircraft and this despite articles highlighting the physical functioning limits of the two techniques. However, we used the information from these early researches to guide our current search for solutions.

III. ORIGINS OF IMAGE REGISTRATION REDUCED PERFORMANCE

A. *Extrinsic physical parameters to the aircraft*

The physical parameters depending on the flight plan of the aircraft contribute in part to the performance of the different image registration. Examples are the speed $v$ of the aircraft, its altitude $h$ of the flight but also the type of topography over which the variation of the richness of its points of interest potentially generates pairings of characteristic points between two distinct images. Point-of-interest matching (or matching descriptors) means the ability to match two points of interest from two separate images with a common overlapping area. The more there are points of interest, the more the result is reliable. Indeed, the error on an image's geolocation according to the application of the algorithm of characteristic points 'descriptors is even lower than there are possible evaluation results to average. If the distribution of the relative positioning error $e$ of a single pair of descriptors into the two images follows a Gaussian distribution law, the mean of the errors $\bar{e}$ is worth 0 with a standard deviation $\sigma$ depending on many physical factors. The calculated displacement of the aircraft has a standard deviation of $\frac{\sigma}{\sqrt{n}}$ where $n$ represents the number of present matching descriptors between the two images. It is therefore better to perform a large number of measurements to improve the geographic positioning accuracy of an image, but it costs time. As we will see, the execution time of a resetting is a determining factor.

Apart from factors external to the aircraft, a number of intrinsic physical parameters play an important role in the performance of the image registration. Among these parameters, for example, we can note the resolution of the optical sensor of the camera but also its longitudinal field of vision $\theta_l$ supposed to scan a dimension of scene parallel to the rectilinear flight movement of the aircraft when the latter is flying in a straight line at altitude $h$. The longitudinal dimension of the scene viewed by the camera is therefore

$$b = 2h \tan\left(\frac{\theta_l}{2}\right).$$

In the case of relative image registration, which is matching common descriptors in the two images, the execution time of a readjustment is limited to the full scan time of the longitudinal dimension of a scene, namely:

$$t_{lim} = \frac{b}{v}$$

The ideal number of descriptor matches between two images should not be chosen randomly. Indeed, the absolute average geographic positioning error called $m$ of an image depends on the standard deviation of the displacement of the aircraft. So, the value of $m$ is proportional to $\frac{1}{\sqrt{n}}$, the execution time of the $t_{exe}$ image registration algorithm is proportional to $n$ with the condition

$$t_{exe} < t_{lim}$$

The figure below gives an example of a parametric graph of $m$ and $t_{exe}$ as a function of $n$.

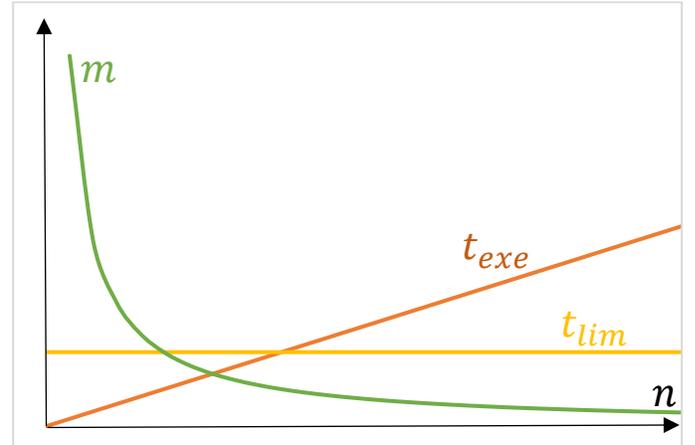

Figure 1: *Evolution of the average error and the execution time according to the calculation load*

At first glance, it is necessary just to find the right compromise between the average error and the execution time of the computing load. The resolution of the sensor has a major role on the average error $m$ and thus on the unitary drift of a simple space registration, whether absolute or relative.

A. *Others cases of performance degradation of readjustment*

Another source of drift, whether for absolute or relative registrations, which degrades the performances of geolocation by image registration technology, lies in the existence of two other situations.

First, like any binary coded iteration, the accuracy of each measurement is limited by the number of bits on which the data is encoded (for example the values of the relative displacements). This error is however, marginal compared to our need in terms of precision.

Another parameter to take into account is the positioning of the pixels on the image. The passage from what is observed (analog perception) to what is stored (digital perception) is not systematically done in the same way, even if we fly over the same landscape. More particularly, when we try to superimpose the pixels, we can see that they are not necessarily superimposed, because the pixel grids are not arranged in the same way. The figure below illustrates the conflict between the matching of identical descriptors when attempting to superimpose transformed images in mathematical sense.

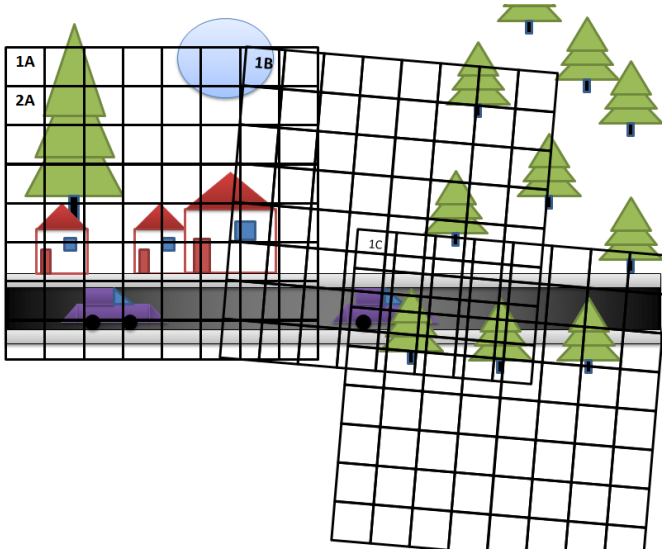

Figure 2 : *Illustration of the impossible superposition between pixels of different shots.*

We clearly see that the pixels of the first two images are straddling each other on the overlapping area. However, two partially overlapping pixels do not exactly cover the same landscape area, and therefore do not have the same value. Hard to imagine making a perfect image registration between them since the two images do not have the same pixels on the area of overlap. In the best case, the superposition would have an error of a half-pixel.

IV. MODELING THE DIFFERENT COUPLING MODES

As mentioned above, a number of non-exhaustive factors cited above degrade the quality of the image registration. We could initially assume that this is a problem of compromise as evoked in Figure 1 above. However, we will show that the mathematical models of the possible coupling modes between the relative and the absolute image registration have precise solutions answering particular needs in terms of physical parametric conditioning.

A. *Introduction of différents coupling modes*

Three modes of coupling are identified in this work. The first mode is the one that is widely used in the various theoretical industrial studies and that we call the *sequential mode*. This operating mode use the relative registration for its speed of calculation and its drift calibrated from time to time by the absolute registration. However, as we will see, this mode can be defeated when it has to run for a long time alternately or redundantly to the GPS system.

The second possible mode of coupling of the image registration, will be called *parallel mode*. This operating mode use the absolute registration in terms of accuracy of measurement but its time of execution is considered too long. So that during the absolute image registration's time of execution, the relative image registration is necessary for the positioning rally, otherwise the flying aircraft has already moved a great distance, which would generate in turn a positioning uncertainty in the form of drift.

Finally, the third mode of possible coupling is the mode which we will call *combined mode* and for which, the performances of the absolute registration make it possible to use the DB embedded with images (sufficiently updated as frequently as possible) to make directly the overlap of images between the current image and an image of the onboard image DB. As a result, this mode makes it possible to use the principle of the relative image registration directly into the embedded image DB.

We will note later and in particular in the following figure RR for Relative image Registration and AR for Absolute image Registration. The figure below summarizes the chronogram of the different types of coupling mode for the image registration process:

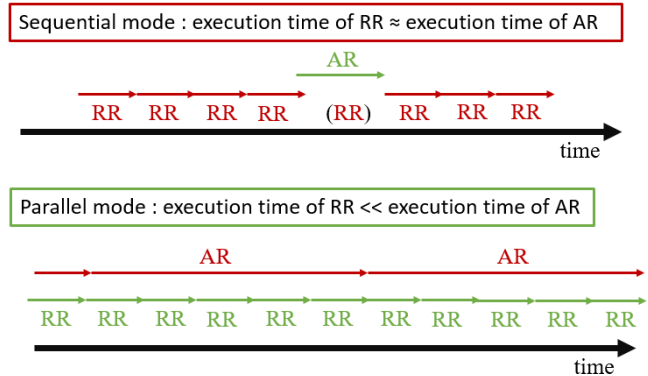

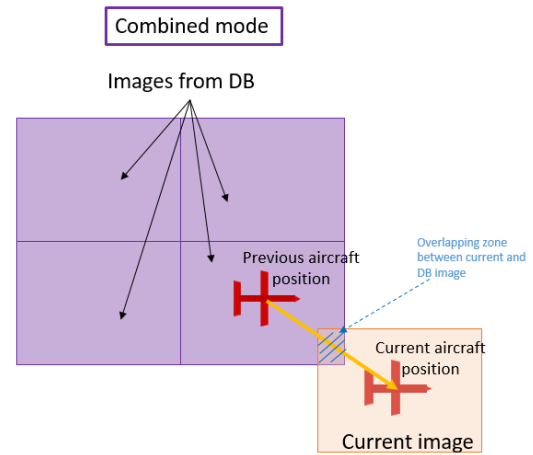

Figure 3 : *Illustration of the different modes of image registration's way of coupling*

*B. Modeling the sequential mode*

The modeling of the sequential mode consists of mathematically formatting the limits of this mode according to the physical characteristics of the system. Absolute and relative image registration have a failure rate, which cannot be considered null. The failure of a realization means that the system has not the ability to provide at a given time; an accuracy result of aircraft's geographic position or even worse is not able to provide squarely the results of a measurement. Even low, this rate, called probability of failure, may not be negligible mostly if plenty of realizations are performed.

The modeling of the sequential mode is based on the possible failure of the absolute image registration with a too long execution time, which not allow system matching a current image with a one from the DB or switching thereafter to a relative image registration process, as showed in Figure 3. Because of this, it is clear that the execution time of the double image registration process in the sequential mode will play a preponderant role in the success of the geolocation measurements.

Noting $p$, the probability of failure of and absolute registration and $N$, the number of times an absolute registration would have been applied during a time $T_0$, the law of large numbers for a sufficiently long time $T_0$ imposes the condition of a not even one failure absolute registration over $T_0$:

$$p \times N(T_0) < 1 \quad (1)$$

The probability of failure $p$ of the absolute registration is even higher than the cumulative drift of the previous relative registration, called $D_T$, would be sufficiently important so that the absolute registration has to explore a critical number of images in its DB, which would delay its execution. Indeed, more complex is the task of the absolute registration to find a particular image in its DB, the more chances of not succeeding are large. We can note that:

$$D_T = n \times D_U \quad (2)$$

With $D_U$, the drift of a single relative image registration and $n$, the number of relative registration between two absolute registrations. We assume that $D_U$ is considered as a constant in our model. It would therefore be important to perform a smaller number $n$ in order to minimize the value of $p$, but during a given time $T_0$, it would maximize the number $N(T_0)$ of absolute image registration. It is no longer a question of compromise but of mathematical optimization.

By noting **a** and **b** image's dimensions of aerial photography, where dimension $b$ point to the direction of the aircraft displacement, the number $\mathcal{N}$ of images to scan in the DB is evaluated as a function of :

$$\mathcal{N} = \frac{(a + I + 2D_T)(b + I + 2D_T)}{ab} \quad (3)$$

Where $I$ represents the supposed inaccuracy absolute image registration process's measurement. The figure below illustrates the concept of drift.

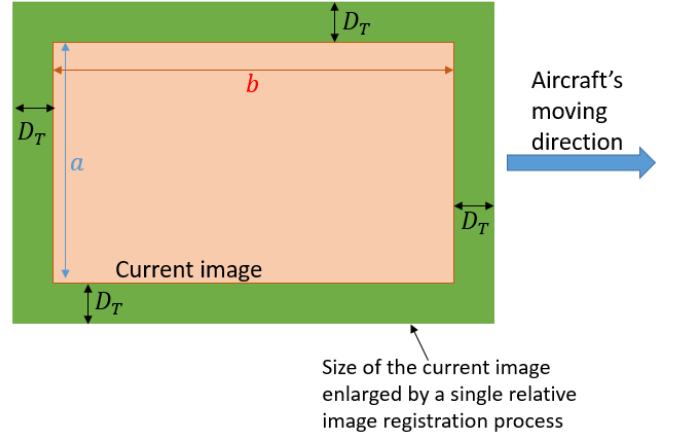

Figure 4 : *Illustration of the cumulative drift for a single relative image registration*

The execution time of an absolute image registration is therefore proportional to the number of images on which potentially, image descriptors can have matching. Using equation (2), and (3), the execution time of an absolute image registration preceded by $n$ relative image registration is:

$$t_{exe} = K_1 \frac{(a + I + 2nD_U)(b + I + 2nD_U)}{ab} \quad (4)$$

With $K_1$ is a proportionality coefficient in second. During $t_{exe}$, the aircraft continues to move for a distance $v \times t_{exe}$. We are modeling the probability of failure of an absolute registration as a function proportional to the execution time $t_{exe}$. Indeed, not only an absolute image registration must permit the next relative image registration to be executed but also, the longer $t_{exe}$ is; the more it would be uncertain for the absolute image registration process to recover the correct current image position or even worse; to be unable to provide any result of current image position. Therefore, we can write:

$$p = K_2 t_{exe} \quad (5)$$

With $K_2$, a proportionality coefficient in $s^{-1}$.

Because any absolute image registration must continue with a relative image registration process, so we have the condition $p = 1$ when $t_{exe} = \frac{b}{v}$. Accordingly, $K_2$ must worth:

$$K_2 = \frac{v}{b} \quad (6)$$

The (5) and (6) involve that the aircraft cannot fly at a speed higher than a certain value by systematically failing the space registration system. This maximum speed is deduced when $=0$ :

$$v_{max} = \frac{ab^2}{K_1(a + I)(b + I)} \quad (7)$$

We can notice that a maximum value of $n$ occurs which systematically induces a failure of the sequential mode, at a given aircraft's speed $v < v_{max}$ such as:

$$n_{max} = \frac{-a - b - 2I + \sqrt{(a-b)^2 + \frac{4ab^2}{K_1 v}}}{4D_U} \quad (8)$$

Outside of these extreme conditions, if we call $t_0$, the execution time of a relative registration, the time of one cycle {n relative + 1 absolute registration} would be:

$$t_{cycle} = nt_0 + K_1 \frac{(a + I + 2n.D_U)(b + I + 2n.D_U)}{ab} \quad (9)$$

Thus, during a $T_0$ aircraft's flight duration under the sequential operating mode, it would have a number $N$ of absolute registration equal to the number of cycles during time $T_0$:

$$N = \frac{T_0}{t_{cycle}} \quad (10)$$

As a result, using equations (4), (7) and (8), we find that $p \times N$ is more like a function of variable $n$ such that:

$$p \times N = f(n) = \frac{T_0 K_1 v \frac{(a + I + 2n.D_U)(b + I + 2n.D_U)}{ab^2}}{nt_0 + K_1 \frac{(a + I + 2n.D_U)(b + I + 2n.D_U)}{ab}}$$

We can simply rewrite

$$f(n) = \frac{T_0 v}{b} \times \frac{[An^2 + Bn + C]}{[An^2 + B'n + C]} \quad (11)$$

With $A = 4D_U^2$, $B = 2D_U(a + b + 2I)$, $C = (a + I)(b + I)$ and $B' = B + \frac{t_0 ab}{K_1}$

By deriving $f(n)$, we obtain

$$\frac{\partial f}{\partial n}(n) = \frac{t_0 T_0 v a}{K_1} \frac{[An^2 - C]}{[An^2 + B'n + C']^2} \quad (12)$$

In order to minimize the number of failures, the optimal number $n$ is then equivalent to:

$$n = \sqrt{\frac{C}{A}} = \frac{\sqrt{(a + I)(b + I)}}{2D_U} \approx \frac{\sqrt{ab}}{2D_U} \quad (13)$$

Provided that $\frac{\sqrt{ab}}{2D_U} \leq n_{max}$, during $T_0$, the smallest value of the average failure number is:

$$p \times N|_{min} \approx \frac{T_0 v}{b} \times \frac{1}{1 + \frac{t_0 ab}{4K_1 D_U \left(\sqrt{ab} + \frac{a+b}{2}\right)}} \quad (14)$$

Therefore, there is a maximum duration of aircraft's flight the chance of failure is high. This condition appears when the condition of the failure number less than 1 is:

$$p \times N|_{min} < 1 \quad (15)$$

Hence, $T_0$ has a maximum value of:

$$T_{0,max} = \frac{b}{v} \times \left[1 + \frac{t_0 ab}{4K_1 D_U \left(\sqrt{ab} + \frac{a+b}{2}\right)}\right] \quad (16)$$

*C. Modeling a parallel mode*

We saw that in the sequential operating mode, the absolute registration must provide a quick result to allow switching on a classic relative registration process. In the parallel operating mode, the absolute registration has not the ability to give the geographic position estimation of the aircraft in a sufficiently short time. As the execution time of an absolute registration is too long, it is necessary in parallel to use relative registration. This operation would permit to rally positions of the aircraft between the moment when the absolute registration has started its calculation and the moment when it finishes its computation and provides the geographic position of the aircraft (estimated position of the aircraft at the start of the calculation of absolute registration). Indeed, the aircraft has fixed wings; thereby it cannot stagnate on "place" the time that the absolute registration succeeds. Relative registration is a tool that allows correcting the future outcome of the current absolute registration process and hence, permitting to the next absolute registration to be done without having to manage too much drift. Modeling the parallel mode involves to evaluating the execution time of the (n+1)-th absolute registration $t_{exe}(n + 1)$ s function of the n-th 's one. The number of relative registration to the n-th absolute registration is $\frac{t_{exe}(n)}{t_0}$ the number of images in the DB is:

$$\mathcal{N}(n) = \frac{(a + I + 2\frac{t_{exe}(n)}{t_0} D_U)(b + I + 2\frac{t_{exe}(n)}{t_0} D_U)}{ab} \quad (17)$$

Thus, there is a recurrence relation such that:

$$t_{exe}(n + 1) = K_1 \frac{(a + I + 2\frac{t_{exe}(n)}{t_0} D_U)(b + I + 2\frac{t_{exe}(n)}{t_0} D_U)}{ab} \quad (18)$$

By simplifying, we get:

$$\Gamma: t_{exe}(n + 1) = \alpha t_{exe}^2(n) + \beta t_{exe}(n) + \gamma \quad (19)$$

With:
$\alpha = \frac{4D_U^2 K_1}{abt_0^2}$ $\beta = \frac{2D_U(a+b+2I)K_1}{abt_0}$ $\gamma = \frac{K_1(a+I)(b+I)}{ab}$

In order for the parallel mode to be viable, it is necessary that the sequence Γ be convergent. It means:

$$(\beta - 1)^2 - 4\alpha\gamma > 0 \quad (20)$$

We set:

$$t_1 = \frac{-(\beta - 1) - \sqrt{(\beta - 1)^2 - 4\alpha\gamma}}{2\alpha} \quad (21)$$

And

$$t_2 = \frac{-(\beta - 1) + \sqrt{(\beta - 1)^2 - 4\alpha\gamma}}{2\alpha} \quad (22)$$

The study of the sequence Γ shows that there is convergence, if $t_{exe}(1)$, the time of the first absolute registration, is between $t_1$ and $t_2$. Or $t_1 \geq 0$ and $t_2 \geq t_1 \geq 0$ if and only if $\beta < 1$. It is thus necessary to perform three conditions for the parallel mode's existence:
Condition 1: $\beta < 1$
Condition 2: $(\beta - 1)^2 - 4\alpha\gamma > 0$
Condition 3: $t_1 \leq t_{exe}(1) \leq t_2$

*D. Modeling the combining mode*

In the combined mode process, the use of absolute registration alone is done not to calibrate any drift but rather to plagiarize the operating mode of a relative registration process between a current image and a DB image. Thus, the relative registration in the sense we have defined it to now no longer exists. Only its concept of overlapping images is used. This mode requires that the images of the DB are those that best match to current overflown ground. Therefore, the DB needs a constant updating to ensure an optimized comparison between a current image and a DB's images stored at an earlier date. The modeling of the combined mode process needs to have the knowledge of the dynamic model of the aircraft for example to evaluate the minimum curvature's radius of the aircraft corresponding to a rotation around a yaw axis. We will call this minimum curvature's radius $R_{min}$. By taking the same notations as for the parallel mode, during the execution time of the n-th relative registration, the aircraft can go straight like using a maneuver of lace with a turning radius between $R_{min}$ and infinite.

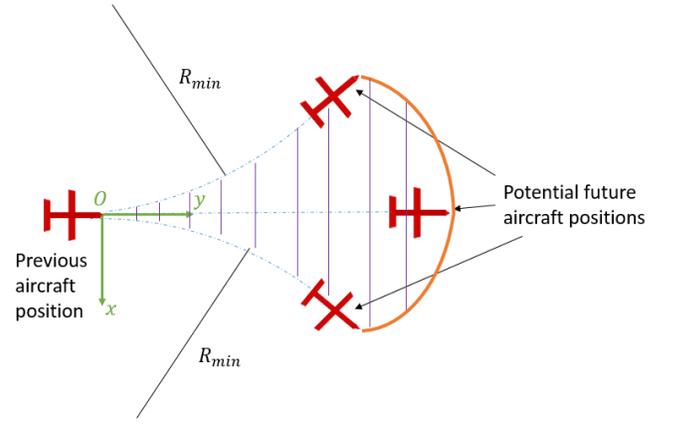

Figure 5 : *Example of possible future aircraft positions during the execution time of an absolute registration according to the combined mode.*

Figure 5 above shows through the illustration of the hatched portion in violet, the uncertainty zone of the aircraft position during the execution time of the previous absolute registration between a current image and DB images. The modeling of the combined mode thus consisted in evaluating the area of this hatched part in figure 5. The greater the value of this area is, the more the next absolute registration will have to exploit images in the DB to find the future position of the aircraft. The coordinate system $(O, x, y)$ as shown in Fig. 5 is introduced as a quantization tool for this area. We evaluate the evolution of the coordinates *of the lower right corner of the image seen by the camera of the aircraft* at any time during the execution time $t_{exe}(n)$ of the previous registration. Therefore, at time $t = 0$, the lower right corner of the image is in O. By noting $R$, the value of an any curvature radius respecting the condition $R_{min} \leq R \leq +\infty$, the parametric limit coordinates of this point are after a given time $t_{exe}(n)$:

$$\begin{cases} x(R) = R\left[1 - \cos\left(\dfrac{v \times t_{exe}(n)}{R}\right)\right] \\ y(R) = R \times \sin\left(\dfrac{v \times t_{exe}(n)}{R}\right) \end{cases} \quad (23)$$

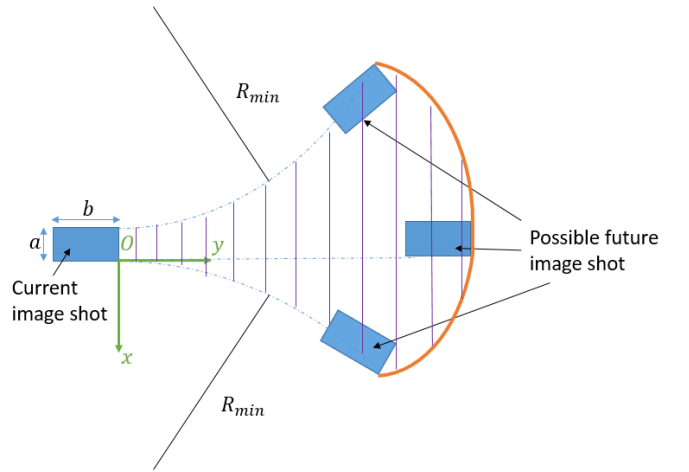

Figure 6 : *Example of possible future image shot positions during the execution time of an absolute registration according to the combined mode.*

As shown in the figure 6 above, for $R_{min} \leq R \leq +\infty$ we obtain the orange curve of Fig. 6 for the part $x \geq 0$, for $-a \leq x \leq 0$, the part of the orange curve is parallel to the axe $(Ox)$. Finally, the part of the orange curve for $-a \geq x$ is the symmetrical part of the orange curve for $x \geq 0$. In our case, we assume that the system is sufficiently powerful to consider the condition:

$$v \times t_{exe}(n) \ll R_{min} \quad (24)$$

It is therefore important for us to evaluate the hatched red area as shown in Fig. 7 below:

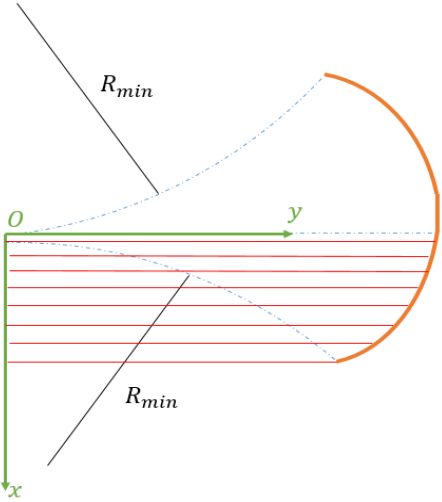

Figure 7 : *Illustration of the hatching of an area of interest*

The hatched area in red above worth:

$$A = \int_{R=+\infty}^{R=R_{min}} y(R) dx(R) \quad (25)$$

The Taylor series of $dx$ leads to the followed approximations:

$$dx(R) \approx -\frac{v^2 \times t_{exe}^2(n)}{2R^2} dR$$

$$y(R) \approx v \times t_{exe}(n) - \frac{v^3 \times t_{exe}^3(n)}{6R^2}$$

Thus, we can evaluate $A$ with approximation:

$$A \approx \frac{v^3 \times t_{exe}^3(n)}{2R_{min}} \quad (26)$$

We deduced that the approximation of the hatched area colored in violet in Fig. 5 for x ≥ 0 worth:

$$\frac{v^3 \times t_{exe}^3(n)}{6R_{min}}$$

As a result, the total hatched area in violet in Fig. 5 corresponding to the aircraft likely zone of presence worth approximately:

$$\Lambda(n) = \frac{2v^3 \times t_{exe}^3(n)}{3R_{min}} + avt_{exe} \quad (27)$$

The number of image (s) to be processed in the DB during the (n + 1)-th absolute registration process is therefore:

$$\mathcal{N}(n) = \frac{\Lambda(n)}{ab} = \frac{2v^3 \times t_{exe}^3(n)}{3R_{min}ab} + \frac{vt_{exe}}{b} \quad (28)$$

This allows us to evaluate the execution time of the (n + 1)- th absolute registration :

$$\Phi: t_{exe}(n+1) = \sigma \times t_{exe}^3(n) + \sigma' \times t_{exe} \quad (29)$$

With

$$\sigma = \frac{2K_1 v^3}{3R_{min}ab}$$

And

$$\sigma' = \frac{K_1 v}{b}$$

We can note that $K_1$ was already defined in the modeling of sequential mode. Let be the equation:

$$E: \sigma x^3 + (\sigma' - 1)x = 0 \quad (30)$$

The series $\Phi$ is convergent if and only if the equation $E$ has 3 real solutions. That means $\sigma' < 1$. If that's the case, the velocity of the aircraft should be less than $\frac{b}{K_1}$ and therefore the 3 solutions of $E$ are $s_1$, $s_2 = 0$ and $s_3$ such as $s_1 = -s_3$ with :

$$s_1 = \sqrt{\frac{1-\sigma'}{\sigma}} \quad (31)$$

The continuation $\Phi$ is convergent if all the conditions quoted below are respected:
Condition 1: $\sigma' < 1$
Condition 2: If $t_{exe}(1)$, that is the time of the first absolute registration, is in the interval $[0, s_1]$.
In this case, $t_{exe}$ converge to 0.

## V. PRACTICAL APPLICATION OF RELATIVE REGISTRATION OF SCENE IMAGES

Beyond the need of modeling the various modes of image registration process, mixing the concept of relative and absolute image registration, it would be useful to provide the mathematical tools that will allow, from the Cartesian coordinates of at least two image descriptors in the common overlap zone, quantifying the evolution of geographic coordinates of the aircraft. The first part of this study will introduce these tools; the second part will show the results of the first performed tests.

### A. *Practical characterization of the movement of the aircraft by relative registration of scene images*

We can ask ourselves what minimal number of image descriptors in common overlap between two images would be necessary to characterize the movement of the aircraft.

We will demonstrate in this section that only two descriptors, in common overlap between two consecutive images, are sufficient to characterize the movement of the aircraft. For this, it is necessary to assign a Cartesian coordinate system to each image, with origins located on the central pixel of each consecutive images. In the figure below, the rectangles represent the consecutive images taken having a common area and are also represented two image descriptors matched between the two shots:

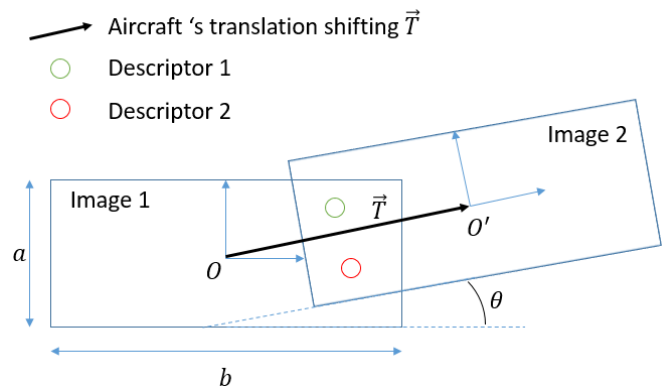

Figure 8 : Illustration of a hypothetic aircraft displacement

The images having a common area will be annotated image 1 and image 2 with respective origins $O$ and $O'$. The movement of the aircraft is entirely characterized by a translation and a rotation. A vector $\vec{T}$ characterizes this translation and an angle $\theta$ characterize the rotation. It is only because the two descriptors called here $D_1$ and $D_2$ will have different coordinates between the two marks that we will be able to characterize the coordinates of the displacement vector $\vec{T}$ in the Cartesian coordinate system $R$ of the image 1 and to estimate the cosine and the sine of the angle $\theta$.

As an input data, the system must be able to evaluate the coordinates of the descriptor $D_1$ as $\binom{x_1}{y_1}$ in the Cartesian coordinate system $R$ of the image 1 and $\binom{x'_1}{y'_1}$ in the Cartesian coordinate system $R'$ of the image 2.

Likewise for the descriptor 2, the system must be able to give us its Cartesian coordinates $\binom{x_2}{y_2}$ in $R$ of the image 1 and $\binom{x'_2}{y'_2}$ in $R'$ of the image 2. Geometrically, the coordinates $\binom{x_1}{y_1}$ of $D_1$ in $R$ becomes $\binom{x'_1}{y'_1}$ in $R'$ by a first rotation of $R'$ relative to $R$ of an angle $\theta$ then by a translation of the vector $\vec{T}$ of coordinates $\binom{t_x}{t_y}$ given in $R$.

Thus, by a simple rotation of an angle $\theta$, we obtain an intermediary reference $R_{int}$ such that the coordinates $\binom{x_1}{y_1}$ of $D_1$ in $R$ become $\binom{x_1 \cos\theta + y_1 \sin\theta}{-x_1 \sin\theta + y_1 \cos\theta}$ in $R_{int}$. After rotating by an angle $\theta$, $y$-Axis and $x$-Axis of $R_{int}$ are parallel to $y$-Axis and $x$-Axis of $R'$. Therefore, there is a simple relation between $t_y$ and $t_x$ : $t_y = t_x \times \tan\theta$

Finally, $R_{int}$ becomes $R'$ by translation $\vec{T}$ with coordinates $\binom{t_x}{t_y}$ given in $R$.

Thus, we can write:
$$\binom{x'_1}{y'_1} = \begin{pmatrix} \cos\theta & \sin\theta \\ -\sin\theta & \cos\theta \end{pmatrix} \times \binom{x_1 - t_x}{y_1 - t_y} \quad (32)$$

In the same way, we have:
$$\binom{x'_2}{y'_2} = \begin{pmatrix} \cos\theta & +\sin\theta \\ -\sin\theta & \cos\theta \end{pmatrix} \times \binom{x_2 - t_x}{y_2 - t_y} \quad (33)$$

We can demonstrate the following relationships:
$$\begin{cases} x_1 = x'_1 \cos\theta - y'_1 \sin\theta + t_x \\ y_1 = x'_1 \sin\theta + y'_1 \cos\theta + t_y \end{cases}$$

$$\begin{cases} x_2 = x'_2 \cos\theta - y'_2 \sin\theta + t_x \\ y_2 = x'_2 \sin\theta + y'_2 \cos\theta + t_y \end{cases}$$

In writing $\Delta x = x_2 - x_1$, $\Delta y = y_2 - y_1$, $\Delta x' = x'_2 - x'_1$ and $\Delta y' = y'_2 - y'_1$, we get rid of the quantities $t_x$ and $t_y$. We obtain the following system:
$$\begin{cases} \Delta x = \Delta x' \cos\theta - \Delta y' \sin\theta \\ \Delta y = \Delta x' \sin\theta + \Delta y' \cos\theta \end{cases}$$
So, we can express value of $\theta$:

$$\cos\theta = \frac{\Delta x' \Delta x + \Delta y' \Delta y}{\Delta x'^2 + \Delta y'^2} \quad (34)$$

$$\sin\theta = \frac{\Delta x' \Delta y - \Delta y' \Delta x}{\Delta x'^2 + \Delta y'^2} \quad (35)$$

We then find the coordinates of the displacement vector as:

$$\begin{cases} t_x = -x'_1 \times \cos\theta + y'_1 \times \sin\theta + x_1 \\ t_y = -x'_1 \times \sin\theta - y'_1 \times \cos\theta + y_1 \end{cases} \quad (36)$$

### B. Use of relative image registration for terrestrial geolocation

The previous chapter allowed us to evaluate the mathematical characteristics of the aircraft's displacement in a local coordinate system related to the acquisition of an original image. Therefore, it is necessary to give the mathematical tools useful to geo position the aircraft according to its geographical coordinates terrestrial (longitude, latitude) based on terrestrial parallels and meridians, shown in the figure below. For the sake of simplicity, this chapter uses the same notations and concepts as the part $A$. above.

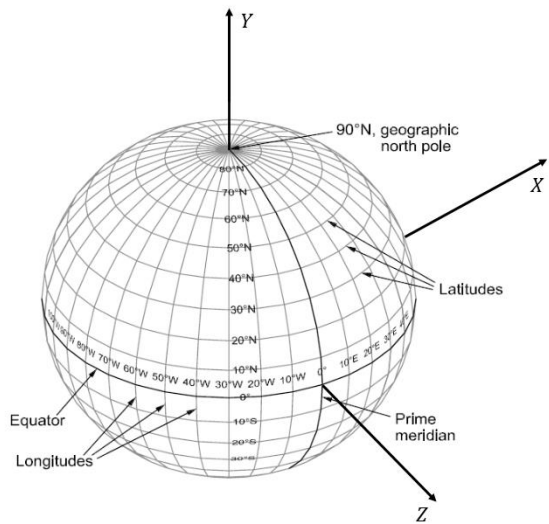

Figure 9 : Geographical coordinates with its Cartesian coordinates system

By noting λ the latitude and φ the longitude (see Figure 9 for the geographic coordinate definitions) of the point $O$, center of the image 1 of Figure 8, the geographical coordinates of the point $O$ are, in the geocentric Cartesian coordinate system $(C, X, Y, Z)$ with its origin at the center of the Earth (supposed to be spherical) named $C$:

$$\overrightarrow{CO} = (R + h) \begin{pmatrix} \cos \lambda \sin \varphi \\ \sin \lambda \\ \cos \lambda \cos \varphi \end{pmatrix} \quad (37)$$

With $R \approx 6370\ Km$, represents the value of the terrestrial radius and $h$, the flight altitude of the aircraft that we consider constant during the maneuver and measured in relation to sea level.
Orientations of $CX, CY$ and $CZ$ axis are given as:
$CX$-axis, oriented from $C$ to $\lambda = 0°$, $\varphi = 90°$ direction
$CY$-axis, oriented from C to $\lambda = 90°$, $\varphi = 0°$ direction
$CZ$-axis, oriented from $C$ to $\lambda = 0°$, $\varphi = 0°$ direction

An angle $\alpha$ exists representing the orientation of the axis $(Ox)$ of the Cartesian coordinates system illustrated on the image 1 in the figure 8, with respect to the $\lambda$-circle of latitude on which the point O is located. See illustration figure 10 below.

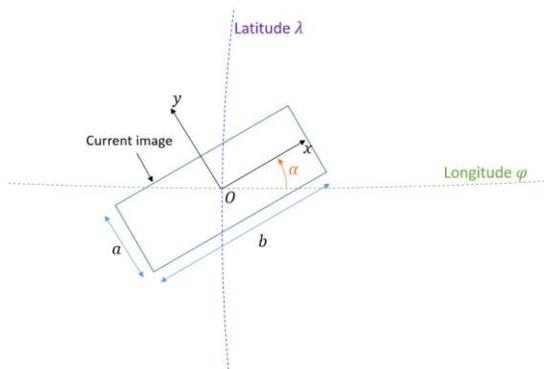

Figure 10 : Orientation of the taken current image compare to the $\lambda$ circle of latitude

By using the same notation as previous Part $A$, the coordinates of the displacement vector $\overrightarrow{T} = (T_X, T_Y, T_Z)$, in the geocentrically Cartesian coordinates system become:

$$\begin{pmatrix} T_X \\ T_Y \\ T_Z \end{pmatrix} = M \times \begin{pmatrix} t_x \\ t_y \end{pmatrix} \quad (38)$$

With

$$M = \begin{pmatrix} \cos \alpha \cos \varphi - \sin \alpha \sin \lambda \sin \varphi & -\sin \alpha \cos \varphi - \cos \alpha \sin \lambda \sin \varphi \\ \sin \alpha \cos \lambda & \cos \alpha \cos \lambda \\ -\cos \alpha \sin \varphi - \sin \alpha \sin \lambda \cos \varphi & \sin \alpha \sin \varphi - \cos \alpha \sin \lambda \cos \varphi \end{pmatrix} \quad (39)$$

By translation of vector $\overrightarrow{T}$, the geocentric coordinates of $O'$, center of image 2 in figure 8, become $\lambda + \Delta\lambda$ for the latitude and $\varphi + \Delta\varphi$ for the longitude, with :

$$\Delta\lambda = \frac{T_Y}{(R + h) \times \cos \lambda} \quad (39)$$

$$\Delta\varphi = \frac{\cos \varphi\, T_X - \sin \varphi\, T_Z}{(R + h) \times \cos \lambda} \quad (40)$$

Moreover, we note by $\alpha'$ the angle equivalent of the angle α defined in Figure 10, physically corresponding to the orientation of the axe $(O'x')$ of the image 2 relative to the $\lambda$ circle of latitude over which the point $O'$ is located. we can demonstrate that there is a relation between $\alpha'$ and α as well as the different physical quantities defined by the cosines and the sinus of the angle $\alpha'$.

$$\begin{cases} \cos \alpha' \approx \cos(\theta + \alpha) + \sin(\theta + \alpha) \sin \lambda\, \Delta\varphi \\ \sin \alpha' \approx \sin(\theta + \alpha) - \cos(\theta + \alpha) \sin \lambda\, \Delta\varphi \end{cases} \quad (41)$$

We see that the value of the angle of the $(O'x')$-axis's with respect to the corresponding circle of latitude passing through $O'$ is close to the value $\theta + \alpha$ corrected by the change of circle of latitude from $\lambda$ to $\lambda + \Delta\lambda$ between the points $O$ et $O'$.

*C. Relatives performances in execution time in relative registration*

If we look closer at the physics of relative image registration, it is uncertain that its process can systematically success in giving a coherent result at the conditions of aircraft flight. In order to be able to perform what is called a relative image registration; algorithms of image matching processing must go through different steps of data processing contained in the 2 images, especially for the need of matching common descriptors. Thus, beyond the image 's loading, the functional analysis of the image processing algorithms makes it possible to highlight the different phases for the data's processing contained in the image allowing to match common descriptors of the two images:

    <u>Phase A:</u> From the set points of interest of the two images, the algorithms must be able to compute the set of the two images' descriptors and match them by pair.

    <u>Phase B:</u> A thresholding process step is a verification

of good matching supposed common descriptors in the two images. It allows removing the matching of two descriptors that would be too different. For each matching, the algorithm looks what is the difference between the two associated descriptors and if "the mathematical distance" between them is greater than a certain threshold then the algorithm delete the matching from the established list of common descriptors.

Phase C: Once the corresponding descriptors have been paired, all that remains is to carry out the step of calculating average translation and rotation for the next image with respect to the preceding image (Calculation of $\vec{T}$ and $\theta$) by using the set of matched pair of descriptors.

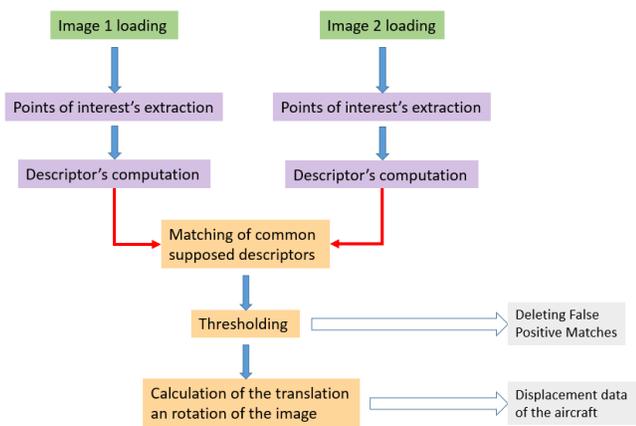

Figure 11 : Calculations process steps for relative registration

The loading phase A of the images depends on the image to be pretreated. This phase includes loading the original image as a pixel matrix and then calculating the corresponding key points and descriptors within the image. The loading time thus comprises a fixed component (for the actual image loading in the form of a matrix) and a variable component depending on the number of descriptors found within the image. By calling $\rho$ the density of descriptors in the image, we can then model our loading phase A time $t_l$ by the following equation:

$$t_l = A_c \times \rho \times ab + L \quad (42)$$

With $A_c$ representing the average time for computing a descriptor of the image and $L$ the time for the actual image loading in the form of a matrix. The images loaded and ready to be compared, the next step will be matching the descriptors of the 2 images. The descriptor matching algorithms take a descriptor of image 1 and randomly compare it to the set of descriptors of the image 2 even if some of them have been already matched in the image 2. Then, the correspondence is made for the descriptors that are the most similar. Either $N$, the average number of descriptors on each image, so $N = \rho ab$, the matching execution time $t_m$ would be:

$$t_m = t_0 N^2 \quad (43)$$

With $t_0$ the execution time for the comparison of a pair of descriptors each one located in one of the two images.
The threshold step is a step of checking the good corresponding of descriptors already matched. Thus, for each false positive (pair of descriptors previously recognized as corresponding but not being in reality), the matching is removed. This step has an execution time independently of the recovery rate of the area of the overlap between the two images. Noting $t_\tau$, the threshold execution time, we can write:

$$t_\tau = B \times N \quad (44)$$

With $B$ the average necessary time to keep or delete a matching. Finally, in the last phase (phase C), it is necessary to establish the calculation of the translation and rotation of image 2 compare to image 1. In order to improve the precision of the calculation of the translation and rotation, the system can give an average value of each of them. The execution time of the calculation is a constant and noted $t_c$.

According to the model and to equations (42), (43) and (44), the total execution time of one relative image registration between two images with an overlap zone, called $T_{exe}$, worth:

$$T_{exe} = t_0 N^2 + (A_c + B) \times N + L + t_c \quad (45)$$

D. *Results on the execution time and the relative registration drift*

We have constituted a test of 72 images (representing an area of 27km²) representing the city of Avignon in France and its surroundings. These 72 images are generally urban-type images, dense in descriptors. We could have worked with a more diverse set of image tests, but the topology of the scene would no longer be urban. Thus, the numerical results presented below are representative only for an urban topology. These 72 images are images of size $512 \times 512$ pixels and an image resolution of 1.19m/pixel. The topography represented is of little importance here, our aim is to have different kind of image panel with a various number of descriptors in order to be able to draw curves of evolutions according to the number of present descriptors. The 72 selected images offer a sufficiently wide panel of descriptors for our experience.

Thus, in order to have a greater diversity of the number of descriptors we have created from 72, 320 possible relative image registration corresponding to 5 types of overlapping percentage (95, 50, 35, 25, 10), that is to say an average of 64 relative registration by percentage overlap to give a tendency to the results 'curves. Therefore, we performed these 320 relative registration and record at each step the execution times of each block of the algorithm. The results below are only averages. The loading time of an image is indeed an increasing linear function of the total number of descriptors in an image.

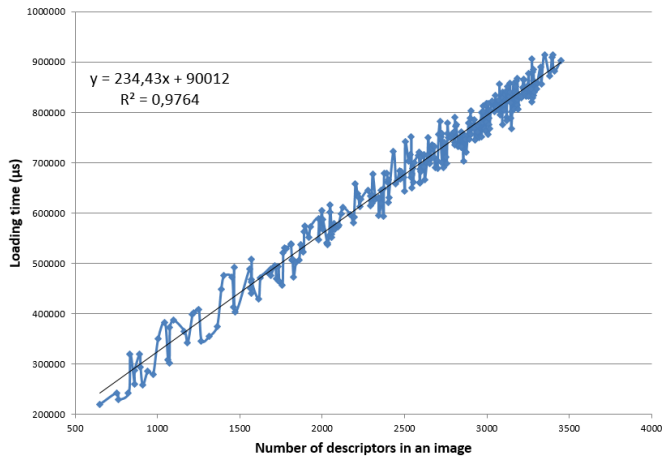

Figure 12 : Evolution of the loading time of an image with its descriptors as a function of the average number of descriptors in an image

Thus, according to equation (42) :
$$A_c \approx 234 \mu s$$
$$L = 9 \mu s$$

Concerning the execution time of matching descriptors given by equation (43), it follows a parabolic function of the second degree of the number of descriptors in an image. The simulation results give the graph below:

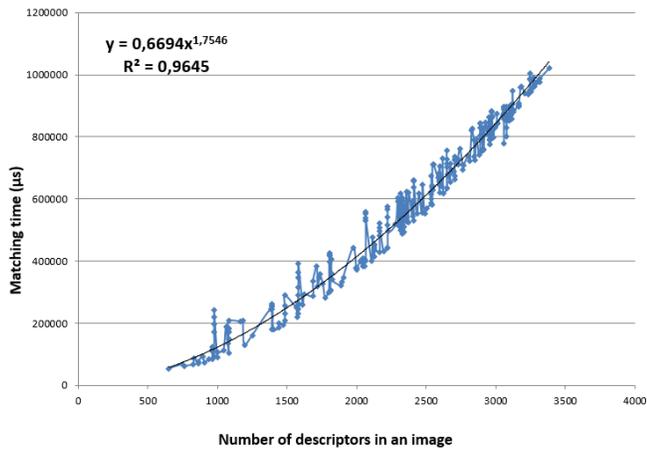

Figure 13 : Evolution of matching descriptor execution time as a function of the average number of descriptors in an image

The shift between the theory and experimentation comes from the various king of topography but also because of computing load of the computer at the time of calculation.
Finally, according to (44) the execution time of the thresholding must evolve as a linear function of number of descriptors in an image.

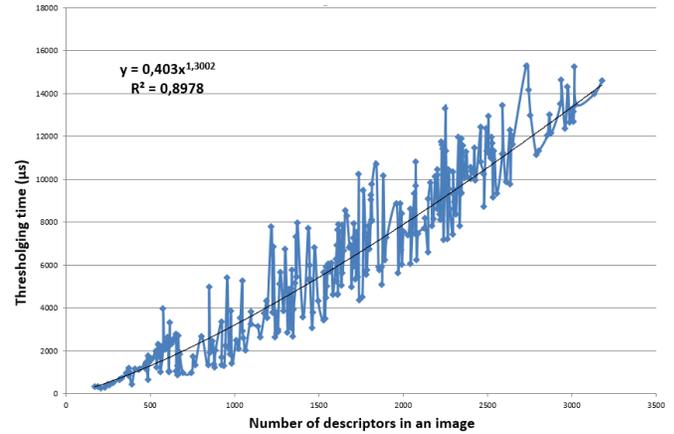

Figure 14 : Evolution of the execution time of the thresholding step as a function of the average number of descriptors in an image

As for the previous case, there is a shift between the theory and experimentation due to experimental conditions.

Concerning the study of the unitary drift of one relative image registration, we obtained the graph below, giving the drift in number of pixels according to the number of descriptors in common and matched for the two images:

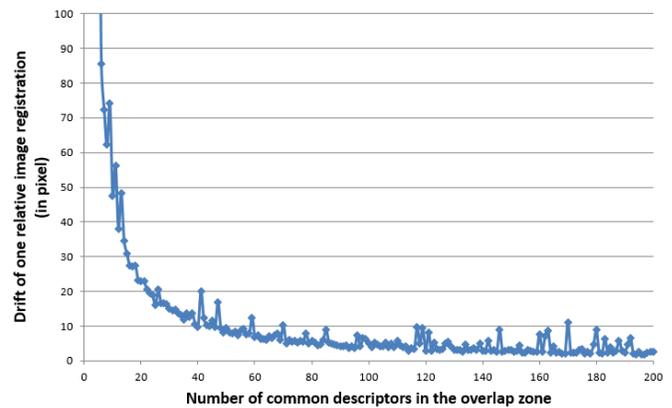

Figure 15 : Evolution of the drift of one relative image registration as a function of the number of common matched descriptors between the two images

We recall that with the chosen resolution of 1.19m/pixel, we see that the unitary drift can be huge if few descriptors are matched. For topographies rich in descriptors, even at the limits, the unitary drift of a relative registration could be limited to less than a meter. However, for environments that are too homogeneous (forest, field of wheat or corn,), as much the performances in execution time of relative registration could be potentially good, as the relative registration's unitary drift could be huge (potentially a hundred meters drift possible). To have less than 1m unitary drift, it is necessary to have more than 600 common matched descriptors in the overlap zone of the two images.

VI. CONCLUSION

Geolocation by image registration is a promising technology capable of replacing inertial units, but this technique requires some restrictions that are evoked either in this article, or at the origin of the modeling assumptions of this study. Thus, the geolocation technique by image registration requires, as for the case of an inertial unit, a systematic and repetitive calibration through an absolute registration using a database embedded in the aircraft. Moreover, this technique requires that the aircraft flies horizontally or be equipped with a system of image stabilizing and rectification in order to avoid creating inaccuracies of measurements on the characterization of each taken aerial image. In addition, the images must be taken in clear weather and especially in daylight, which is an important constraint.

Regarding relative registration performance, which is a major advantage for this king of geolocation technique, which permits to reduce the execution time of the previous technique of registration based on the use of image database, and they relate only to the topology overflown. In this article, models of the three different ways of linking the two types of image registration were elaborated without being put to the test of experimentations.

Nevertheless, the first studies of relative registration sensitivities give us a trend of how relative registration behaves. For homogeneous environments (forest type, sea or mountain), few descriptors would describe the images, which would cause a great inaccuracy on the positioning of the next image due to a large drift but, this has for asset however to reduce the execution time of a relative registration. On the contrary, for heterogeneous topographies, many descriptors in common between two images taken consecutively would make it possible to drop significantly the drift in the next image positioning calculation, but this will have the disadvantage of creating a high run time of the relative registration, which would potentially jeopardize the next one.

We thus see that the technique of geolocation by image registration is a complex technique to master with the entanglement of many physical parameters, which act according to their value, on the performances of this technique and this according to the type of topography overflown.

A future technique could use artificial intelligence that would be led, according to the topography overflown, to choose the parameters of the algorithms but also the physical parameters of the aircraft such as speed or altitude in order to go toward optimal performance of this technique and this, for each picture taken by the ventral camera embedded on the aircraft.

ACKNOWLEDGMENT

The authors would like to thank Lea GJATA for her devotion to the stimulating innovation center's team.

REFERENCES

[1] Arrêté du 11 avril 2012 relatif à la conception des aéronefs civils qui circulent sans aucune personne à bord, aux conditions de leur emploi et sur les capacités requises des personnes qui les utilisent. NOR: DEVA1206042A. Version consolidée au 11 mars 2015

[2] Basic Principles of Inertial Navigation, site des étudiants en aéronautique de l'université de Delft

[3] I. LEBLOND, M. LEGRIS, B. Solaiman. « Apport de la classification automatique d'images sonar pour le recalage à long terme, » 2008.

[4] A. ANSAR and L. MATTHIES, "Multi-modal image registration for localization in Titan's atmosphere," In *Proceedings of the 2009 IEEE/RSJ international conference on Intelligent robots and systems*. IEEE Press, p. 3349-3354, 2009.

[5] V. KÜTEKIN, « *Navigation and control studies on cruise missiles*. » Thèse de doctorat. MIDDLE EAST TECHNICAL UNIVERSITY 2007.

[6] E. ROYER « Cartographie 3D et localisation par vision monoculaire pour la navigation autonome d'un robot mobile. ». Thèse de doctorat. Université Blaise Pascal-Clermont-Ferrand II, 2006.

[7] M. J. ROUX LOPEZ-KRAHE, H. MAÎTRE. « Recalage image SPOT/carte routière. » *International Archives of Photogrammetry and Remote Sensing*, vol. 29, p. 384-384, 1993.

[8] G. CONTE, M. P. HEMPELRUDOL and *al,* "High accuracy ground target geo-location using autonomous micro aerial vehicle platforms." In *Proceedings of the AIAA-08 Guidance, Navigation, and Control Conference*. 2008.

[9] M. H. P. R. D. L. S. D. Gianpaolo Conte, «High Accuracy Ground Target Geo-location Using Autonomous Micro Aerial Vehicle Platforms,» 2008.

[10] A. CHAO, M. BURKE, T. KURIEN, L. CICO, "*Real-Time Geo-Registration on High-Performance Computers*." Defense Technical Information Center, 2002.

[11] K. M. HAN, G. N . DESOUZA, "Geolocation of multiple targets from airborne video without terrain data," *Journal of Intelligent & Robotic Systems*, vol. 62, no 1, p. 159-183, 2011.

[12] R.-H. P. S. M. I. R.-C. K. M. I. Dong-Gyu Sim, «Integrated Position Estimation Using Aerial Image Sequences,» 2002.

[13] [12] G. C. a. P. Doherty, «An Integrated UAV Navigation System Based on Aerial Image Matching,» 2077.

[14] K. m. H. a. G. DeSouza, «Geolocation of Multiple Targets from Airborne video without terrain Data,» 2008.